\documentclass{epl}

\title{Averaged residence times of  stochastic
motions in bounded domains}
\author{O. B{\' e}nichou\inst{1} \and M.
Coppey\inst{1} \and M.
Moreau\inst{1} \and P.H. Suet\inst{1} \and R.
Voituriez\inst{2}  }
\institute{
  \inst{1} Laboratoire de physique th\'eorique des
liquides - 4, place
Jussieu
75255 Paris Cedex 05 France\\
  \inst{2} Institut Curie - 26 rue d'Ulm 75248 Paris cedex 05 France
}

\begin{document}

\maketitle

\begin{abstract}
Last year, Blanco and Fournier (Blanco S. and Fournier R., Europhys. Lett. 2003) calculated
 the mean first exit time
of a domain of a particle undergoing a randomly reoriented ballistic motion which starts from the boundary. They showed that it is   simply
related to the ratio of the volume's domain over its surface. This work was extended by Mazzolo (Mazzolo A., Europhys. Lett. 2004) who studied the
case of trajectories which start inside the volume. In this letter, we propose an alternative  formulation of the problem which allows us to
calculate not only the mean exit time, but also the  mean residence time inside a sub-domain. The cases of any combinations of reflecting and
absorbing boundary conditions are considered. Lastly, we generalize our results for a wide class of stochastic motions.
\end{abstract}

\section{Introduction}

Recently, motivated by  the study of diffusive trajectories performed  by animals in various conditions,  Blanco and Fournier \cite{blanco} reported an important result concerning the mean first
exit time of Pearson random walks \cite{hugues} in a bounded domain. Beyond the applications for animal trajectories, numerous physical systems
are concerned, such as the neutron scattering processes \cite{doering,freund}. Pearson random walks can be defined by the trajectory of a particle
which is submitted at stochastic times to random reorientations  of the direction of its constant velocity $v$. The stochastic times are
exponentially distributed with mean value $1/\lambda$. For this category of random walks, Blanco and Fournier showed that the mean first exit time
$<t>$ of a random walk starting from the boundary of a finite domain is independent of the frequency $\lambda$ and is simply related to the ratio
of the domain's volume $V$ over the surface $S$ of the domain's boundary, namely in three dimension:

\begin{eqnarray} \label{blanco}
<t>=\frac{4}{v}\frac{V}{S}
\end{eqnarray}
 The fact that this result does not depend on the
frequency $\lambda$ but only on the geometry of the system seems counter-intuitive, although Blanco and Fournier propose some heuristic
explanation for it. Very recently, following the demonstration of  Blanco and Fournier, Mazzolo \cite{mazzolo1} extended their results and   gave
a relation between the $n$th moment of the first exit time of a particle starting inside the domain, and the $n+1$th moment of the first exit time
of a particle starting from the boundary.

In this paper, we present an alternative approach, which allows us to extend these results in three directions. First, we consider the case of any
combination of absorbing and reflecting boundary conditions, which is useful in many applied stochastic processes. Second, we study not only first
exit time properties, but also residence time within a sub-domain properties (prior to the first exit of the domain), which is a more general
quantity. Last,
 we
show that these results can be generalized for a wide class of stochastic motions.

\section{First exit time}

Before considering the case of a general motion, we restrict our discussion to the previously examined Pearson walks, which corresponds to a
succession of deterministic ballistic movements in the bounded domain $V$ delimited by an absorbing boundary $\Sigma$, interrupted by
instantaneous and isotropic redistributions of the velocity $\vec{v}$. Let $p(\vec{r'},\vec{v'},t|\vec{r},\vec{v})$ be the conditional density
probability at time $t$ that the particle is  at position $\vec{r'}$ with a velocity $\vec{v'}$, given that it starts initially at position
$\vec{r}$ with a velocity $\vec{v}$. This quantity satisfies the well-known backward Chapman-Kolmogorov differential equation \cite{gardiner}:

\begin{eqnarray} \label{bckde}
\displaystyle
\partial_{t}p(\vec{r'},\vec{v'},t|\vec{r},\vec{v})=\vec{v}\cdot\overrightarrow{\nabla}_{\vec{r}}p(\vec{r'},\vec{v'},t|\vec{r},\vec{v})+\frac{\lambda}{\sigma_{d}}\int
\mbox{d}\tilde{v''} [p(\vec{r'},\vec{v'},t|\vec{r},\vec{v''})-p(\vec{r'},\vec{v'},t|\vec{r},\vec{v})],
\end{eqnarray}
where
$\sigma_{d}=\frac{2\pi^{\frac{d}{2}}}{\Gamma(\frac{d}{2})}$
stands for
the solid angle in $d$-dimension and the integral $\int {\rm d} {\tilde v}$ is taken over unit vectors in the direction of the velocity. This equation can be
easily
converted into an equation for the moments of the
 first
exit time $T(\vec{r},\vec{v})$ starting from $\vec{r}$ with the initial velocity $\vec{v}$\cite{risken}:
\begin{eqnarray} \label{timemoment}
\displaystyle \vec{v} \cdot
{\overrightarrow{\nabla}_{\vec{r}}}t_{n}(\vec{r},\vec{v})+\frac{\lambda}{\sigma_{d}}\int
d \tilde{v''}
(t_{n}(\vec{r},\vec{v''})-t_{n}(\vec{r},\vec{v}))=-nt_{n-1}(\vec{r},\vec{v}),
\end{eqnarray}
where $t_n(\vec{r},\vec{v})$ is the $n$th moment of the first exit time, which verifies $t_n(\vec{r},\vec{v})=0$ on the absorbing boundary
($\vec{r}\in\Sigma$), for all $n\geq 1$ and for an outward velocity. Noting that the symmetric quantity $\int{\mbox{d}\tilde{v} \mbox{d}\tilde{v''}
t_{n}(\vec{r},\vec{v''})-\int \mbox{d}\tilde{v} \mbox{d}\tilde{v''} t_{n}(\vec{r},\vec{v})}$ obviously equals zero, and that $\vec{v} \cdot
{\overrightarrow{\nabla}_{\vec{r}}}t_{n}(\vec{r},\vec{v})=\mbox{div}(t_{n}(\vec{r},\vec{v})\vec{v})$, the integration of equation
(\ref{timemoment}) over all possible initial positions and velocities gives:
\begin{eqnarray} \label{intt}
\displaystyle \int \mbox{d}\tilde{v} \int_{V} \mbox{d}\vec{r} \mbox{div}(t_{n}(\vec{r},\vec{v})\vec{v})=-n \int \mbox{d}\tilde{v}\int_{V}
\mbox{d}\vec{r} t_{n-1}(\vec{r},\vec{v}).
\end{eqnarray}
Applying the Gauss divergence theorem on the left-hand side of the equation (\ref{intt}), we can write that:
\begin{eqnarray} \label{timeSV}
\displaystyle
<t_{n}>_{\Sigma}=\eta_d \frac{n}{v} \frac{V}{\Sigma}<t_{n-1}>_{V},
\end{eqnarray}
where the two averages involved in equation (\ref{timeSV}) are defined for any function $f$ as follows :
\begin{eqnarray} \label{moy}
<f>_{\Sigma}=-\frac{1}{v\Sigma\alpha_{d}} \int \mbox{d}\tilde{v} \int_{\Sigma} \mbox{d}\vec{\Sigma} \cdot \vec{v} f(\vec{r},\vec{v}), \mbox{ and }
<f>_{V}=\frac{1}{V\sigma_{d}} \int \mbox{d}\tilde{v} \int_{V} \mbox{d}\vec{r} f(\vec{r},\vec{v})
\end{eqnarray}
with $\displaystyle \alpha_{d}=\frac{2\pi^{\frac{d-1}{2}}}{(d-1)\Gamma(\frac{d-1}{2})}$ is the inward flux of a unit, isotropically distributed
vector through a unit surface, and $\displaystyle \eta_d=\frac{\sigma_{d}}{\alpha_d}=\sqrt{\pi}
(d-1)\frac{\Gamma\left(\frac{d-1}{2}\right)}{\Gamma\left(\frac{d}{2}\right)}$ is a dimension dependent constant.
Equation (\ref{timeSV}) can be rewritten as:
\begin{eqnarray} \label{fmat}
\displaystyle
<t_{1}>_{\Sigma}=\frac{\eta_d}{v}\frac{V}{\Sigma}, \mbox{ and } \forall n\geq 1,  \;
<t_{n-1}>_{V}=\frac{<t_{n}>_{\Sigma}}{n<t_{1}>_{\Sigma}}
\end{eqnarray}
which corresponds to the results previously obtained by Blanco and Fournier\cite{blanco} and Mazzolo\cite{mazzolo1}. Parenthetically, we know that
since $<t_{2}>_{\Sigma}\geq<t_{1}>^{2}_{\Sigma}$, (\ref{fmat}) yields a general lower bound for $<t_{1}>_{V}$:
\begin{eqnarray}
\displaystyle
<t_{1}>_{V}\geq\frac{1}{2}<t_{1}>_{\Sigma}=\frac{\eta_d}{2v}\frac{V}{\Sigma}.
\end{eqnarray}

\section{General boundary conditions}

\begin{figure}
\twofigures[scale=0.3]{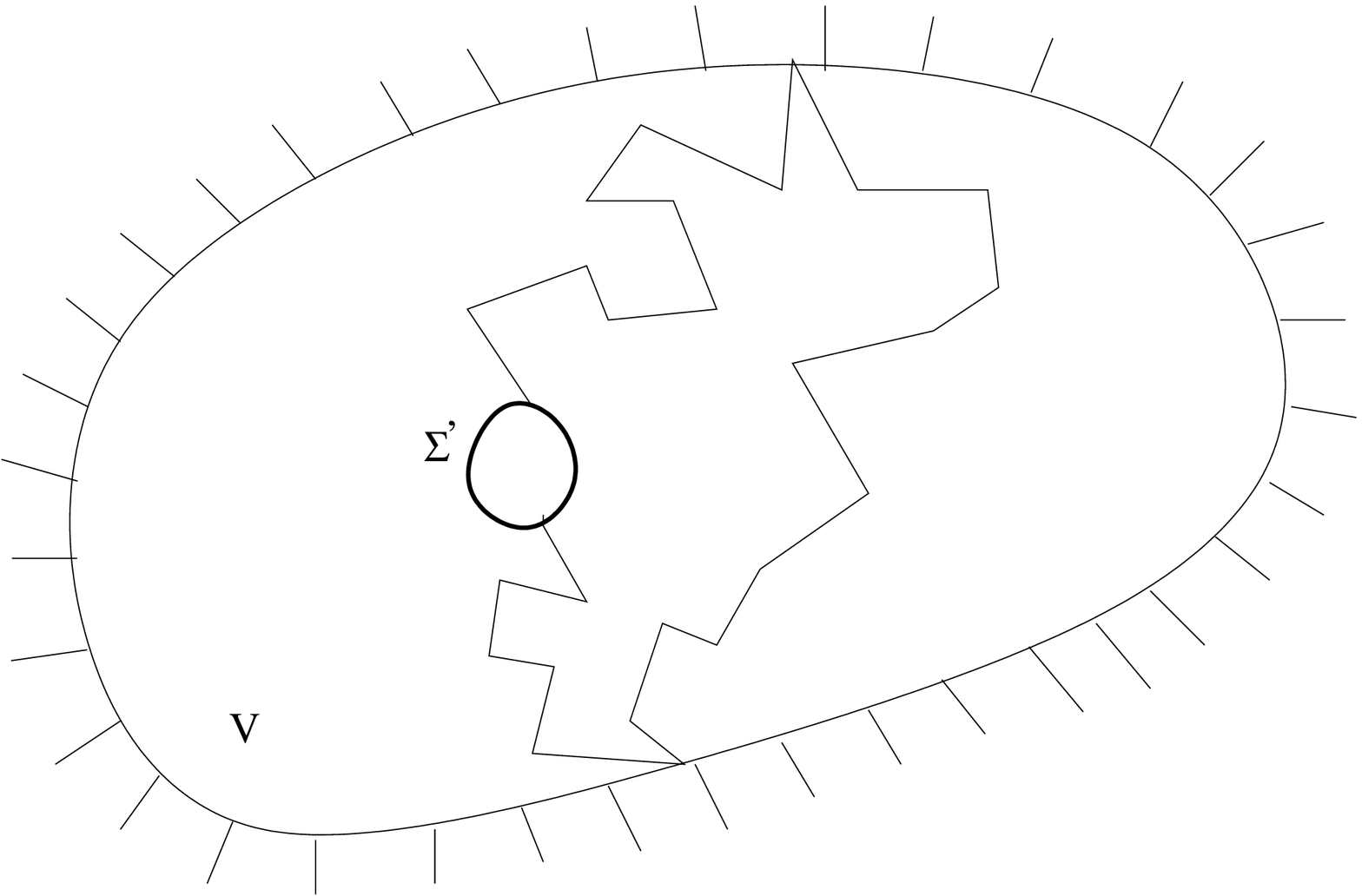}{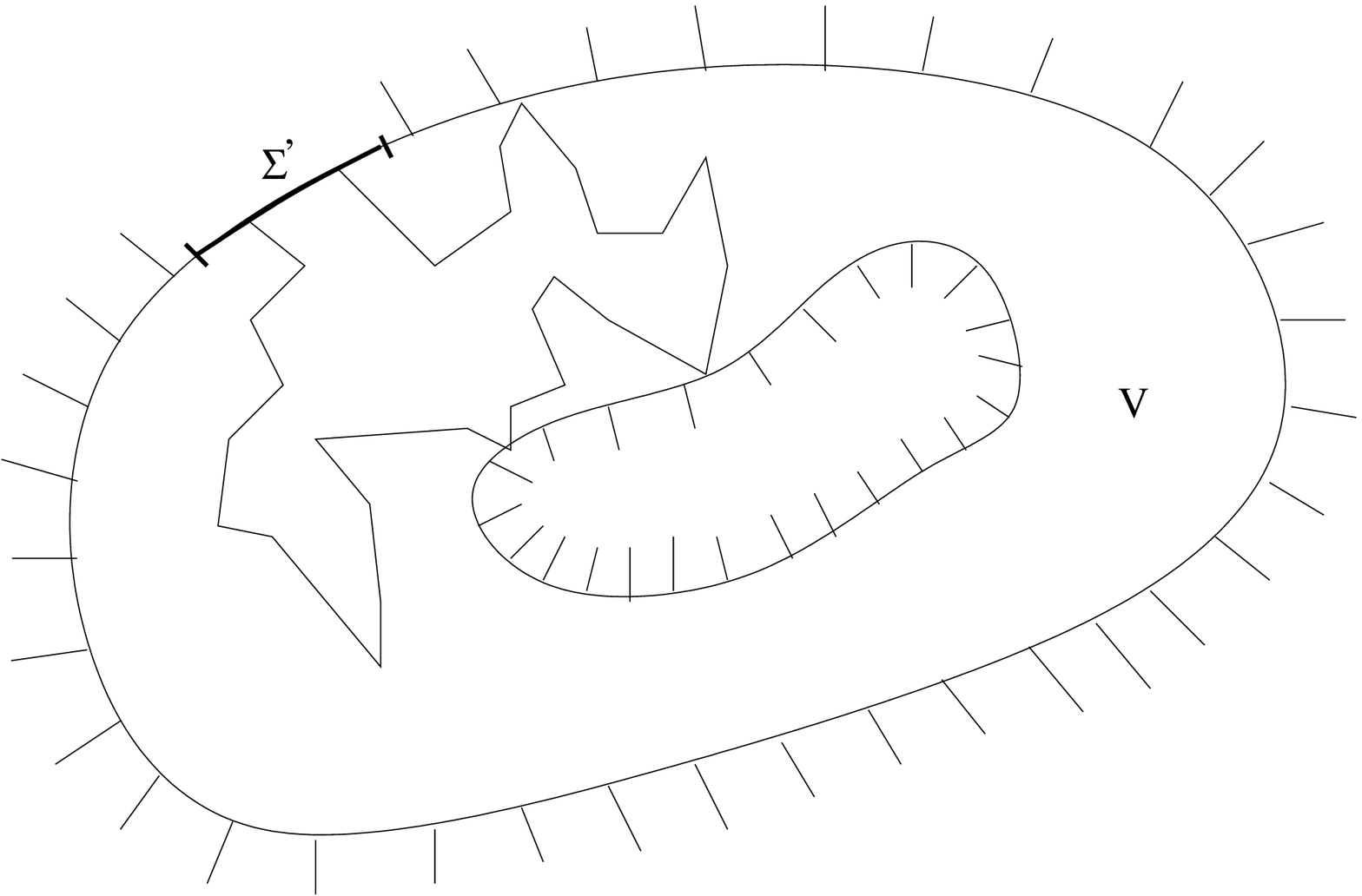} \caption{A particular particle's trajectory which starts and ends on the absorbing boundary
$\Sigma_{\mbox{abs}}$, for an absorbing region inside a closed reflecting surface $\Sigma$} \label{f.1} \caption{A particular particle's
trajectory for a reflecting box with a hole (aperture) $\Sigma_{\mbox{abs}}$} \label{f.2}
\end{figure}
We now extend these  results, which take unexpected simple forms, to the general case of mixed boundary conditions. The domain's boundary $\Sigma$
is now supposed to be composed of any combination of absorbing and reflecting parts (see fig.\ref{f.1},\ref{f.2}). Let $\Sigma_{\mbox{abs}}$ be
the absorbing part of $\Sigma$, and $\Sigma_{\mbox{refl}}$ its reflecting part, so that $\Sigma=\Sigma_{\mbox{abs}}+\Sigma_{\mbox{refl}}$. We
assume that initially the particle starts from the absorbing surface.
Whether we consider that the reflecting boundary is perfect, i.e. that the inward velocity angle equals the outward velocity angle, or whether we
consider that the particle's velocity angle is uniformly redistributed after each collision with the reflecting boundary, we have
\begin{eqnarray} \label{reflect}
\displaystyle \int \mbox{d}\tilde{v}\int_{\Sigma_{\mbox{refl}}} \mbox{d}\vec{\Sigma}\cdot \vec{v} t_{n}(\vec{r},\vec{v})=0.
\end{eqnarray}
 In this case of mixed boundary conditions, equation
(\ref{fmat}) becomes :
\begin{eqnarray} \label{fmatgeneral}
\displaystyle <t_{1}>_{\Sigma_{\mbox{abs}}}=\frac{\eta_d}{v}\frac{V}{\Sigma_{\mbox{abs}}}, \mbox{ and } \forall n\geq 1,  \;
<t_{n-1}>_{V}=\frac{<t_{n}>_{\Sigma_{\mbox{abs}}}}{n<t_{1}>_{\Sigma_{\mbox{abs}}}}
\end{eqnarray}

Let us, for instance, consider animals leaving their home  through its boundary $\Sigma_{\mbox{abs}}$, and exploring a domain $V$ (see
Fig.\ref{f.1}). The previous equation gives a simple estimate of their mean return time to home if the boundary $\Sigma_{\mbox{refl}}$ is
reflecting for them, provided that the interrupted ballistic motion correctly models their behaviour.

\section{Splitting probabilities}

Consider now a closed volume $V$ with two absorbing surfaces $\Sigma_{1}$ and $\Sigma_{2}$, and let us compute the absorption probability
$\Pi_1(\vec{r},\vec{v})$ on the surface $\Sigma_{1}$ for a particle starting initially from the boundary $\Sigma_{\mbox{abs}}=\Sigma_1+\Sigma_2$.
This probability
 obeys the differential equation \cite{van}:
\begin{eqnarray}
\displaystyle \vec{v}\cdot\overrightarrow{\nabla}_{\vec{r}}  \; \Pi_1(\vec{r},\vec{v})+\frac{\lambda}{\sigma_{d}}\int \mbox{d}\tilde{v''}
[\Pi_1(\vec{r},\vec{v''})-\Pi_1(\vec{r},\vec{v})]=0,
\end{eqnarray}
for which the boundary conditions, for an outward velocity $\vec{v}$, are $\Pi_1(\vec{r},\vec{v})_{\vec{r}\in\Sigma_{1}}=1$ and $\displaystyle
\Pi_1(\vec{r},\vec{v})_{\vec{r}\in\Sigma_{2}}=0$. Following the same lines as previously, one can show that
\begin{eqnarray} \label{split}
\displaystyle <\Pi_1>_{\Sigma_{\mbox{abs}}}=\frac{\Sigma_{1}}{\Sigma_{1}+\Sigma_{2}}.
\end{eqnarray}
In other words, the splitting probability through absorbing boundary is uniform, when starting uniformly from any point of it.

\section{Residence time}

\begin{figure}
\onefigure[scale=0.3]{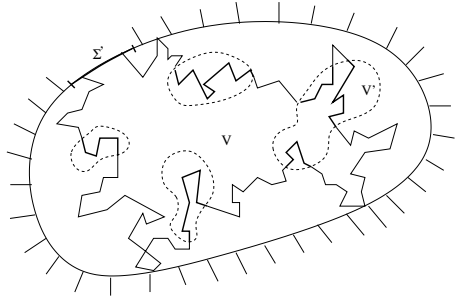} \caption{Residence time of the particle inside a sub-domain of the volume.} \label{f.3}
\end{figure}

A relevant quantity involved in the theory of chemical reactivity and in other similar problems\cite{bere,wf,agmon,join}, is the residence time in
a sub-domain $V'$ (see fig.\ref{f.3}) of the domain $V$. In order to study it, we assume that the particle can disappear in $V'$ (but not outside
$V'$) with a constant and uniform reaction rate $k$. The central quantity is now the
 survival probability of the particle with
respect to the
reaction in  volume $V'$ (i.e. considering that {\it an absorption on the
boundary is not a
killing process}), which can be expressed in
two
ways \cite{kac,bere,weiss}:
\begin{eqnarray}\label{defS}
\displaystyle S(t|\vec{r},\vec{v})&=&\int \mbox{d}\vec{r'} \mbox{d}\tilde{v'} p(\vec{r'},\vec{v'},t|\vec{r},\vec{v}) =<e^{-kT_{t}(\vec{r},\vec{v})}>
\end{eqnarray}
where $T_{t}$ is the residence time up to the observation time $t$ inside $V'$ of the particle,  starting from $(\vec{r},\vec{v})$, and
$p(\vec{r'},\vec{v'},t|\vec{r},\vec{v})$ is the conditional density probability of the particle
 which satisfies
the
following backward Chapman-Kolmogorov differential
equation:

\begin{eqnarray} \label{bck}
\displaystyle \nonumber
\partial_{t}p(\vec{r'},\vec{v'},t|\vec{r},\vec{v})&=&\vec{v}\cdot\overrightarrow{\nabla}_{\vec{r}}p(\vec{r'},\vec{v'},t|\vec{r},\vec{v})+\frac{\lambda}{\sigma_d}\int
\mbox{d}\tilde{v''}
[p(\vec{r'},\vec{v'},t|\vec{r},\vec{v''})-p(\vec{r'},\vec{v'},t|\vec{r},\vec{v})]\\
&&-k{\bf 1} _{V'}(\vec{r})
p(\vec{r'},\vec{v'},t|\vec{r},\vec{v})
\end{eqnarray}
where ${\bf 1} _{V'}(\vec{r})$ is the indicator
function of $V'$ which equals
zero if $\vec{r}\notin V'$ and equals one if
$\vec{r}\in V'$. Note that
the boundary conditions
give $S(t|\vec{r},\vec{v})=1$ for an outward velocity of a particle on the absorbing surface $\Sigma_{\mbox{abs}}$. Indeed, such a particle
absorbed on this surface will never disappear within the sub-domain $V'$. Writing down the equation (\ref{bck}) for the conditional survival
probability, we obtain:
\begin{eqnarray}\label{Seq}
\displaystyle
\partial_{t}S(t|\vec{r},\vec{v}) & = & \vec{v}\cdot
\overrightarrow{\nabla}_{\vec{r}}S(t|\vec{r},\vec{v})+\frac{\lambda}{\sigma_d}\int \mbox{d}\tilde{v''}
[S(t|\vec{r},\vec{v''})-S(t|\vec{r},\vec{v})]\nonumber\\
                                   &   &  -k {\bf
1}_{V'}(\vec{r}) S(t|\vec{r},\vec{v}).
\end{eqnarray}
As $S(t|\vec{r},\vec{v})$ is a bounded non-increasing function of $t$, it tends to a finite limit as $t\to\infty$, so much $\displaystyle
\frac{\partial S}{\partial t}\to 0$. Letting $t\to\infty$ in (\ref{Seq}), integrating it over $\vec{r}\in V$ and $\vec{v}$, taking into account
the boundary conditions and using the expansion (see (\ref{defS}) when $t\to\infty$)
\begin{eqnarray} \label{expan}
\displaystyle S_{\infty}(\vec{r},\vec{v})=\sum_{n=0}^{+\infty}(-1)^{n}\frac{k^{n}}{n!}\tau_{n}(\vec{r},\vec{v})
\end{eqnarray}
where $\tau_{n}(\vec{r},\vec{v})$ is the $n$th moment of the residence time of the particle inside $V'$ for an infinite observation time, we
derive our main result  for the residence time:
\begin{eqnarray} \label{residencetime}
<\tau_{1}>_{\Sigma_{\mbox{abs}}}=\frac{\eta_{d}}{v}\frac{V'}{\Sigma_{\mbox{abs}}}
 \;\;{\rm and}\;\;\forall n \geq 1, \;
<\tau_{n-1}>_{V'}=\frac{<\tau_{n}>_{\Sigma_{\mbox{abs}}}}{n<\tau_{1}>_{\Sigma_{\mbox{abs}}}}.
\end{eqnarray}


Note that the  mean fraction of time spent inside $V'$ prior to the first exit of $V$ is given by
$<\tau_{1}>_{\Sigma_{\mbox{abs}}}/<t_{1}>_{\Sigma_{\mbox{abs}}}=V'/V$, which can be seen as an ergodic type property\cite{van}.

\section{Generalisation}

Let us now assume that the particle undergoes a more general stochastic diffusion process, interrupted at stochastic times by an instantaneous
redistribution of its velocity, without changing its position. The conditional density probability $p(\vec{r'},\vec{v'},t|\vec{r},\vec{v})$ to be
at position $\vec{r'}$ with velocity $\vec{v'}$ at time $t$, starting from $(\vec{r},\vec{v})$ at time 0, corresponding to the general diffusion
process obeys the backward equation
\begin{equation}\label{general}
\partial_t p= v_i\partial_{x_i} p + \frac{F_i}{m}
\partial_{v_i}
p+D_{ij}\partial^{2}_{v_iv_j} p \equiv {\rm L}^+p
\end{equation}
where $m$ is the mass of the particle, $\vec{F}(\vec{r},\vec{v})$ is the force exerted on the particle in state $(\vec{r},\vec{v})$, $D_{ij}$ is
the general diffusion matrix, and $L^+$ stands for the adjoint of $L$. In many situations of physical interest, $\vec{F}$ is the sum of a
conservative force due to a potential $U$, $-\partial_{\vec{r}} U$, and of a friction force $-\eta \vec{v}$ due to a surrounding fluid, $\eta$
being a constant friction coefficient. Furthermore, $D_{ij}=\delta_{ij}D\eta ^2/m^{2}$, $D$ being the usual diffusion coefficient of the Fick law,
which is independent of $\vec{r}$ in an homogeneous medium. Now, we assume that this general diffusion process is interrupted at stochastic times,
for which the velocity $\vec {v}$ of  the particle is instantaneously redistributed with a given transition rate
$q(\vec{r},\vec{v'}|\vec{r},\vec{v})$.  The overall stochastic process, including the velocity reorientation,
obeys the equation:
\begin{eqnarray}\label{general2}
\partial_t p(\vec{r'},\vec{v'},t|\vec{r},\vec{v})&=&
{\rm L}^+p(\vec{r'},\vec{v'},t|\vec{r},\vec{v})+ \int {\rm
d}\vec{v''}q(\vec{r},\vec{v''}|\vec{r},\vec{v})[(p(\vec{r'},\vec{v'},t|\vec{r},\vec{v''})-
p(\vec{r'},\vec{v'},t|\vec{r},\vec{v}))] \nonumber\\
&\equiv&{\cal L}^{+} p(\vec{r'},\vec{v'},t|\vec{r},\vec{v}).
\end{eqnarray}
The conditional probability density at time $t$, $p(\vec{r'},\vec{v'},t|\vec{r},\vec{v})$ also obeys the following forward equation:
\begin{equation}
\partial_t
p(\vec{r'},\vec{v'},t|\vec{r},\vec{v})={\cal L}p(\vec{r'},\vec{v'},t|\vec{r},\vec{v}).
\end{equation}
Let us assume that this equation, when the whole boundary $\Sigma$ is assumed to be reflecting, admits a stationary solution
$p_0(\vec{r},\vec{v})$. For instance $p_0(\vec{r},\vec{v})$ is the thermodynamic equilibrium $\propto e^{-(U+\frac{mv^{2}}{2})/k_BT}$ if equation
(\ref{general}) is the usual diffusion equation of a particle in an equilibrium fluid at temperature $T$ and if the velocity redistribution
process defined previously does not affect this thermodynamic equilibrium, which is the case if  $\int {\rm d} {\vec v'} 
(q(\vec{r},\vec{v'}|\vec{r},\vec{v})p_0(\vec{r},\vec{v})-q(\vec{r},\vec{v}|\vec{r},\vec{v'})p_0( \vec{r},\vec{v'}))=0$. Then it is possible to generalize the previous results. Indeed, let us
consider the residence time of the particle in a sub-domain $V'\subset V$, and let $\Sigma_{\mbox{abs}}$ be the absorbing part of the total
surface $\Sigma$. As previously, we assume that inside $V'$, the particle can disappear with a constant and uniform reaction rate $k$. Its
conditional survival probability at time $t$, $S(t|\vec{r},\vec{v})=\int \mbox{d}\vec{r'}\mbox{d}\vec{v'}p(\vec{r'},\vec{v'},t|\vec{r},\vec{v})$,
starting from $(\vec{r}\in V,\vec{v})$ obeys the backward equation
\begin{eqnarray} \label{38}
\displaystyle \frac{\partial  }{\partial t} S(t|\vec{r},\vec{v})= {\cal L}^{+} S(t|\vec{r},\vec{v})-k {\bf 1} _{V'}(\vec{r}) S(t|\vec{r},\vec{v})
\end{eqnarray}
where ${\bf 1}_{V'}(\vec{r})=1 $ if $\vec{r}\in V'$ and 0 otherwise. Following the method used for the Pearson random walks,  we let $t\to+\infty$
in (\ref{38}), and we multiply this equation by the equilibrium probability $p_{0}(\vec{r},\vec{v})$ computed for the reflecting conditions on the
boundary $\Sigma$ of $V$. The integration of equation (\ref{38}) over all $\vec{r}\in V$ and $\vec{v}$, leads to:
\begin{eqnarray}\label{20}
\displaystyle \int_{\vec{r}\in V} \mbox{d}\vec{r} \int \mbox{d}\vec{v}p_{0}(\vec{r},\vec{v}) {\cal
L}^{+}S_{\infty}(\vec{r},\vec{v})=k\int_{\vec{r}\in V'}\mbox{d}\vec{r} \int \mbox{d}\vec{v} p_{0}(\vec{r},\vec{v}) S_{\infty}(\vec{r},\vec{v}).
\end{eqnarray}
The left hand side of (\ref{20}) can be
transformed into

\begin{eqnarray}
\displaystyle \int_{\vec{r}\in V} \mbox{d}\vec{r} \int\mbox{d}\vec{v} S_{\infty}(\vec{r},\vec{v}) {\cal L} p_{0}(\vec{r},\vec{v})
+\int_{\vec{r}\in \Sigma} \mbox{d}\vec{\Sigma} \cdot\int \mbox{d}{\vec v} \vec{v} S_{\infty}(\vec{r},\vec{v}) p_{0}(\vec{r},\vec{v}).
\end{eqnarray}
The first integral vanishes, since $ {\cal L}p_0(\vec{r},\vec{v})=0$ if $\vec{r} \in V$. In the second integral we separate the inward and outward
integration over $\vec{v}$ in the surface integral.
 The inward part can be written, using the expansion (\ref{expan}) of $S_\infty$: $\displaystyle \sum_{n=0}^\infty (-1)^{n+1}\frac{k^n}{n!}J<\tau_{n}>_{\Sigma_{\mbox{abs}}}$, where
$J=-\int_{\vec{r}\in \Sigma_{\mbox{abs}}} \mbox{d}\vec{\Sigma} \cdot \int \mbox{d}\vec{v} \vec{v}  p_{0}(\vec{r},\vec{v})$
 is the one-way equilibrium
probability current on $\Sigma_{\mbox{abs}}$, and the  surface average of a quantity $f(\vec{r},\vec{v})$ is defined as:
\begin{equation}
<f>_{\Sigma_{\mbox{abs}}}=-\frac{1}{J}\int_{\vec{r}\in\Sigma_{\mbox{abs}}}d{\vec \Sigma}\cdot\int_{\vec{v}_{in}} \mbox{d}\vec{v} \vec{v}f(\vec{r},\vec{v})p_0(\vec{r},\vec{v}).
\end{equation}
The outward part is actually equal to  $J$, as $S_\infty(\vec{r},\vec{v})=1$ if $\vec{ r}\in\Sigma$ and $\vec{v}$ points outward.   
 Then, comparing both sides of (\ref{20}), 
  we generalize the relations (\ref{residencetime}) for the moments of the residence time within $V'$:
\begin{eqnarray}
\displaystyle <\tau_{1}>_{\Sigma_{\mbox{abs}}}=\frac{P}{J}\;\;\mbox{ and }\;\; <\tau_{n}>_{\Sigma_{\mbox{abs}}}=n\frac{<\tau_{n-1}>_{V'}}{J}
\end{eqnarray}
where $P=\int_{\vec{r}\in V'} \mbox{d}\vec{r} \mbox{d}\vec{v} p_{0}(\vec{r},\vec{v})$ is the equilibrium probability of volume $V'$.

Considering now the special case $V'=V$, we obtain the relations for the moments  of the first exit time  from $V$:
\begin{eqnarray} \label{genetps}
<t_{1}>_{\Sigma_{\mbox{abs}}}=\frac{1}{J}, \; \; \mbox{ and } \; \;
<t_{n-1}>_{V}=\frac{1}{n}\frac{<t_{n}>_{\Sigma_{\mbox{abs}}}}{<t_{1}>_{\Sigma_{\mbox{abs}}}}.
\end{eqnarray}
which generalize the formulas (\ref{fmat}). In the same way, the results of the splitting probabilities (\ref{split}) can be generalized for the
general stochastic motion discussed here. Equation (\ref{genetps}) takes a particularly simple form if we now consider a fluid of identical
particles and if we replace $p_0(\vec{r},\vec{v})$ by the equilibrium density of particles in the phase space $(\vec{r},\vec{v})$. Then, the
probability current $J$ is replaced by the equilibrium one-way flux of particles $\phi$ through $\Sigma$, and equation (\ref{genetps}) becomes:

\begin{equation}
<t_{1}>_{\Sigma}=\frac{N}{\phi}
\end{equation}
$N$ being the number of particle in $V$. This formula can easily be understood intuitively: if at time $0$ the boundary is supposed to be
reflecting, the number of particles leaving $V$ per unit time should be $\sim\frac{1}{<t_{1}>_{\Sigma}}$, so that the  equilibrium is maintained
inside $V$ if $V$ is supplied with an entering flux of  new particles $\phi\sim\frac{N}{<t_{1}>_{\Sigma}}$.

\section{Conclusion}

We have shown that the geometrical relations previously obtained for Person random walks in bounded domains, are particular cases of very general
relations between residence times for a large class of stochastic processes. Thus, they can be very useful in many applications, when the
evolution equations can not be solved exactly. Furthermore they can be extended to intermittent systems, which are frequent in
biology\cite{benichou}. Lastly, with relevant changes, this work can be extended to discrete space and time systems. Such  extensions are in progress.

\acknowledgments
We thank Sylvain Condamin for useful discussions.

\end{document}